\documentclass[aps,prd,twocolumn,groupedaddress,nofootinbib]{revtex4}

\usepackage{graphicx}
\usepackage{amsmath,amssymb,amsfonts}
\usepackage[colorlinks,citecolor=blue,linkcolor=blue,urlcolor=blue]{hyperref}
\usepackage[applemac]{inputenc}

\usepackage{physics} 
\usepackage[percent]{overpic} 

\newcommand{\be}{\begin{equation}}
\newcommand{\ee}{\end{equation}}
\usepackage[english]{babel}

\usepackage[autostyle, english = british]{csquotes}
\MakeOuterQuote{"}

\begin{document}

\title{Agency in Physics}

\author{Carlo Rovelli}
\affiliation{Aix Marseille University, Universit\'e de Toulon, CNRS, CPT, 13288 Marseille, France.\\ Perimeter Institute, 31 Caroline Street North, Waterloo, Ontario, Canada, N2L 2Y5.\\ The Rotman Institute of Philosophy, 1151 Richmond St.~N London, Ontario, Canada, N6A 5B7.}

\date{\small\today}

\begin{abstract} \noindent   I discuss three aspects of the notion of agency from the standpoint of physics: (i) what makes a physical system an agent; (ii)  the reason for agency's time orientation; (iii)  the source of the information generated in choosing an action.  I observe that agency is  the breaking of an approximation under which dynamics appears closed.   I distinguish different notions of agency, and observe that the answer to the questions above differ in different cases.  I notice a structural similarity between agency and  memory, that allows us to model agency, trace its time asymmetry to thermodynamical irreversibility, and identify the source of the information generated by agency in the growth of entropy.  Agency is therefore a physical mechanism that transforms low entropy into information.  This may be the general mechanism at the source of the whole information on which biology builds. 

 \end{abstract}

\maketitle

\section{\bf The problem} 

Agency is the possibility for an agent to \emph{act} on the world, and affect it.  The notion of agency is used in a variety of contexts, with variable meanings.  Agents play a role in areas spacing from economy to theology.  They are increasingly utilised in foundational contexts, for instance in discussing the conceptual basis of thermodynamics \cite{Maxwell1871,Myrvold}, quantum mechanics \cite{Fuchs2019}, causality \cite{Ramsey1978,Price2007,Pearl2000,Woodward2016}, even the foundations of physics itself \cite{Deutsch2013b}. 

Agency raises three questions for a physicist.  First, how to understand the assumed independence of the agent and its possibility of alternative choices, given that real agents are themselves physical systems that do not violate laws of nature.  Second, agency affects the future, not the past: what is the origin of the time asymmetry, considering that the elementary laws of physics are invariant under time reversal? Third, an agent can choose alternative courses of action and the choice generates information: an agent can pick one among $N$ alternatives, generating $I=\log_2 N$ bits of information. Where does this information originate from? 

Here I consider a solutions to these three questions. On various other recent physical perspectives on agency, see \cite{Kirchhoff2018,Kolchinsky2018,Still2020,Durham2020} and references therein. A  perspective similar to the one considered here has been independently developed very recently by Barry Loewer in \cite{Loewer2020}.  I give for understood that nothing in agency conflicts with known laws of nature; but understanding how the actual behaviour that we denote agency can be accounted for in terms of these laws is something that requires a bit of thinking.  This is what is done here.  

To this end, I consider a general characterisation of agency, but also distinguish distinct manners in which the notion of agency is intended and used. These capture different degrees of the independence, or freedom, we attribute to the agent. 

To account for agency's time asymmetry we cannot recur to the time orientation of the agent's perspective (as is done in many contexts), because this is what we want to account for, not to assume.  The only viable alternative is to trace it to the manifest time-asymmetry of the macroscopic world.   This, in turn, is accounted for by the second principle of thermodynamics, widely understood; by which I mean here the genericity assumption of statistical physics plus (the non-genericity assumption of) the past hypothesis \cite{Albert2000}, namely the fact that entropy  was low in the past.  The time orientation of agency must be ultimately rooted in the second principle because there is nothing else at our disposal  \cite{Price,Albert2015}.  As we shall see, however, this is realised indirectly and in different ways for different kinds of agencies. The core of the paper is Section \ref{Sec:4}, where I show how  the dots between thermodynamics and agency can be filled-in, using a simple model that illustrates how entropy growth can drive agency.  

The model presented here derives from a structural similarity between agency and memory, and a recent results on the relation between memory and entropy \cite{Rovelli2020}. In particular, the model yields a thermodynamical bound on the information  produced in agency, tying the generation of information to thermodynamical parameters.    

Memory and agency can thus be viewed as mechanisms that convert free energy into information. This may well be the primary source of the information the biosphere, the brain, and culture, deal with.

\section{\bf The Agent as a physical system} 

The key to address the nature of agency, is to recognise that agency does not refer solely to events in the world: it refers to a manner of description of these events. The notion of agency is grounded in ignoring physical links, namely some of the physical (deterministic or probabilistic) correlations described by the physical laws.

\subsection{\bf Agency is disregarding physical links} 

To illustrate this idea, consider the use of agency in the foundations of thermodynamics or quantum theory.   A  fertile formulation of both of these theories is to see them describing the response of  physical systems (thermodynamical or quantum) when they are \emph{acted upon} in certain manners. For instance: "If we compress the volume of a gas, the temperature increases so and so". Or:  "If we prepare a q-bit in this state, and then measure this spin, we obtain this number".   These are descriptions of behaviours of a part of the world, when an \emph{agent} acts on it one way or the other.   The language of agency is explicit in numerous presentations of these theories, and is sometimes deemed essential. 

A moment of reflection, however, shows that this language can be translated away.  Any occurrence of "if an agent acts on the system in this and that manner" can be translated into a statement of the form "if the system happens to interact in this and that manner"; thus trading the independence of the agent with the modality that is at the basis of all physical laws.   

Physical laws, indeed, refer to regularities, namely to repetitive behaviour happening under repeated circumstances. They are generically of the form "Anytime that A then also B", or "anytime that A then the probability of B is so and so". The "anytime" is a conditional ("if").  The phase space of classical mechanics and the Hilbert space of quantum theory are spaces of possibilities, where the conditionals reside.  Laws have been found, in principle, by generalisation and induction out of a number of \emph{repeated} observations.   Hence the notion of an agent "free to act" is actually irrelevant in the foundations of thermodynamics and quantum theory: it can be replaced by the conditional: "whenever this, then that". 

But the opposite is equally true. And it is more interesting. Precisely because physics is modal in this manner, we can always replace the conditionals with the action of an independent agent.  And express the arbitrariness by attributing it to something that we call "agency".   Therefore the agent is here simply the place where we arbitrarily decide to start the sequence of correlations described by the laws we are interested in: it is, in other words, where we ignore previous physical links. 

To illustrate this, consider for instance the statement that the temperature of a mass of real gas increases when compressed.  The compression is due to the interaction between the gas and some other physical system.  This other physical system can be a human agent freely deciding to push a piston; but also the wind pushing a mass of atmospheric air downhill along a mountain.  For the gas, which is what is being considered, the difference is irrelevant: the human and the wind are "agents". What makes them agents, here, is simply the fact that in describing the behaviour of the gas we are not interested in the chain of physical links they might happen to follow: these are treated as external, arbitrary.  It is this that makes them agents here: ignoring their physics. 

This is in fact general.  Agency is always associated to the boundaries of an incomplete or approximate description of the world, within which physical links are to some extent closed, namely approximately sufficient to account for the evolution.  It refers to the spots where the physical links are ignored. The agent is the system whose physical links are neglected in a given account.  To see how this works in general, however, we have to distinguish variants of the notion of agency.

\subsection{Different notions of agency}

 In a wide sense, any physical system acting on a second system and affecting it can be called an agent. But the word "agency" is commonly employed in a more restricted sense, indicating the capacity of certain systems, such as humans, to take independent, autonomous, intentional decisions and act on these.   

The  ambiguity in the use of the term is reflected in the philosophical debate about agency (see for instance \cite{Schlosser2019} and references there).  From the perspective of physics, the ambiguity refers to the assumptions about the reasons for an agent to act in one way or the other.  There is a spectrum of (overlapping) possibilities, leading to distinct notions of agency, which can be denoted as follows. We can call: 
\begin{description}
\itemsep-.3mm 
\item{\em External agent:} any system, when we simply disregard the reasons for its behaviour. Example: in dealing with the dynamics of the Moon's surface, a meteorite that impacts on its surface is an external agent. 
\item{\em Internal agent:} a system governed by some complex internal dynamics which we could reconstruct. Example: This computer is the agent that controls that door.  
\item{\em Random agent:} a system governed by a genuinely probabilistic dynamics.  
\item{\em Independent agent:} a system governed by an internal (deterministic or probabilistic) dynamics, too complex for us to reconstruct. Example: This man is the agent that decides whether to open that door.  
\item{\em Supernatural agent:} a system that does not satisfy neither deterministic nor probabilistic physical laws.   
\end{description}
\emph{External} agency is only a way of talking about external physical links when we are not interested in accounting for them. \emph{Random} agency can be instantiated by quantum theory. Human agency is an example of \emph{independent} agency \cite{Spinoza1677b,Ismael2016}. The existence of agency that does not to satisfy neither deterministic nor probabilistic physical laws (\emph{supernatural}) would contradict our current understanding of our world I see no interest in considering it.  The most interesting case is \emph{independent} agency, in particular when the agent can choose between alternatives that affect the world differently. In Section \ref{Sec:4} we shall see how a physical system can actually do so.

 \section{\bf Time orientation} 

Agency is time oriented: it affects the future, not the past.  What is the source of this time asymmetry?   The answer is delicate, because it differs for different notions of agency \cite{Price}. 

 \subsection{\bf Perspectival time orientation} 

Let's start with the simplest case. Consider an elastic collision between a ball ${\cal B}_1$ and a ball ${\cal B}_2$.  When it is hit by  ${\cal B}_1$, the ball ${\cal B}_2$ changes its velocity.  Say the velocity before the collision was $\vec v_{past}$  and after the collision it becomes $\vec v_{future}$.  We can say that ${\cal B}_1$ has acted on ${\cal B}_2$ and the effect of this action is in the future: it has changed $\vec v_{past}$ into $\vec v_{future}$.   This is a possible example of an action affecting the future. 

However, the physical laws governing the collision are time reversible.  There is nothing in the process itself that picks up a time direction.   At given past $\vec v_{past}$ , it is the future to be affected by the act; but at given future $\vec v_{future}$, it is the past to be affected. That is: at fixed past, the world with the collision and the world without the collision have a different future; while at fixed future, the world with the collision and the world without the collision have a different past.   We could equally describe the same history backward in time, with the same laws, and say that the effect of the interaction has been to change the velocity from $\vec v_{future}$ to $\vec v_{past}$.

The reason we say that the collision affects the trajectory of the particle ${\cal B}_2$ \emph{after} the collision is only to be found  in the regard \emph{we} give to the phenomenon. It is \emph{we} who are time oriented. In turn, the reason \emph{we} take the past as fixed is that we can remember it and we cannot influence it, while we cannot remember the future and we can influence it.  Hence we consider the past states of the two balls as given, and we say that the effect of the collision is in the future.  The distinction refers to what \emph{we} know, not to anything in the phenomenon itself.  The distinction is perspectival.  As far the phenomenon alone is concerned, it is purely linguistic: we simply \emph{call} effect what happens \emph{after} the collision \cite{Hume1736}.  

It is tempting to jump from this to saying that this is all there is to say about the time orientation of agency: it is perspectival, agency looks time oriented, but it is only because we see it so.    But that would be a mistake. 

The reason is that we have simply displaced the problem: the collision does not distinguish  cause froms effect, but \emph{we} do.   And \emph{our} distinction is rooted in our own agency, which can affect the future but not the past.  The phenomena determined by \emph{us} and \emph{our} agency ---and with us a large class of other systems we call internal agents--- are definitely not time symmetric.

In particular, to have a different effect on the ball ${\cal B}_2$, a \emph{different} motion of the ball ${\cal B}_1$ is needed, while I can now choose between different macroscopic futures given the  \emph{same} macroscopic past.  What is the source of time orientation in \emph{this} case?  

 \subsection{\bf Physical time orientation} 

It is not difficult to find the source of time-oriented phenomena: the entire macroscopic world around us is manifestly time oriented.   We understand this time orientation of the macro-world in terms of  the second principle (in a generalised sense, and including the past hypothesis) which is the only "fundamental" law that breaks time-reversal invariance.  There is no reason for agency to be different, and there is no other source of time orientation available in our universe (see below for a discussion about quantum theory).  
Agency must therefore be a macroscopic phenomenon governed by an entropy gradient (and ultimately the past hypothesis of a primordial low entropy that underpins it)  \cite{Price,Albert2015}.  This must be the ground for the orientation of complex agents like us.   

This is the only possible answer to the question of the origin of the time orientation of agency, in the context of a naturalistic perspective. The main question I address in this paper is \emph{how} an entropy gradient can give rise to the behaviour we recognise as agency.  As we shall see in the next Section, the additional ingredients needed for this are surprisingly meagre. 

The subtle point is the fact that it is the \emph{macroscopic} world to be time oriented. The micro-history of reality happens to be such that in a direction of time (the "past") the microstate belonged to a low entropy macrostate. (To even state time orientation we need to have a notion of macrostates, namely a coarse graining.)  Accordingly, agency must be accounted form in terms of a macroscopic/microscopic distinction, in the sense of statistical mechanics and  thermodynamics.  

This is not a distinction on the basis of size, scale, or number of degrees of freedom; it is a distinction relative to a set variables (called macroscopic), to which we have access and that have a partially closed dynamics within some approximation. That is,  their behaviour can be approximately determined without involving other variables. (Here I take low initial entropy, or the past hypothesis, as given: I do not discuss the possibility for itself to be perspectival, which is discussed in \cite{Rovelli:2015}.)    

Now, consider an internal agent. If we described it in complete mechanical terms, the time orientation of agency would  again be just a linguistic choice.   But, as we have seen in the previous Section, an agent is precisely a system of which we are disregarding part the dynamics.  When we describe a human being as an agent, we \emph{are} obviously not describing its complete microphysics.  Hence, the separation between manifest (macroscopic) degrees of freedom and underlying (microscopic) ones that are not accounted for is constitutive to the notion of independent agency \cite{Spinoza1677b}. It is precisely this separation that underpins the thermodynamical roots of agency's time orientation. 

The general situation is therefore clear: the root of time orientation in an independent agent is thermodynamical irreversibility.  This underpins independent agency, hence our own sense of openness of the future. This, in turn, gives us the perspective to read even trivial symmetric interaction in a time oriented manner. 
 
What is missing is to unravel a mechanism showing how the thermodynamical irreversibility can account for the time orientation of agency and the openness of the future it implies.  This is what is done in the next section. 

\section{Modelling the thermodynamical irreversibility of agency} \label{Sec:4}

Consider an independent agent: a complex unpredictable macroscopic system.  Say that in the interval between the times $t_a$ and $t_b$ it acts on the macroscopic world causing an effect.  Say it can choose between $N$ alternatives in its action.  Consider the time evolution of the macroscopic state of the world, including the agent itself, and call it $Q_i(t)$ with $i=1,...,N$ labelling the $N$ possible evolutions (or "branches") of the macro-world. The branches have the same history before the action and differ after-wise, that is
\begin{eqnarray}
Q_i(t)&=&Q_j(t), \ {\rm for\ all}\ i,j\ {\rm and  }\ t<t_a,\nonumber\\
Q_i(t)&\ne&Q_j(t),  \  {\rm for\ some}\ i,j\ {\rm and  }\  t>t_b.  
\label{agency}
\end{eqnarray}
This describes what an independent agent, capable of choosing, does. 

The internal dynamics of the agent can be a complex computation about possible futures, based on the memory and a value system incorporated in the agent's memory or structure (more on this below); it can be a random process influenced by the indeterminism of quantum mechanics (more on this below), or by microscopic statistical fluctuations; or it can simply be any classical dynamics too complex for us to reconstruct. The relative weight of these components in the indeterminacy of the macroscopic evolution is irrelevant from the point of view of physics, because in all cases it simply amounts to disregarding some physical links in the evolution. 

Let's disregard for the moment quantum indeterminism.  We picture the situation as follows: a macroscopic deterministic dynamics gives a good approximation to the dynamics of each $Q_i(t)$ for any $t$, but not in the interval  $t_a-t_b$ during which agency acts.

The key point is that this is not in contradiction with classical determinism, because there is a large number of micro-histories $q(t)$ compatible with anyone of the branches of the evolution in \eqref{agency}.  Hence there is nothing mysterious in the branching itself: it is just a case where the causal closure of the (approximate!) macroscopic dynamics breaks down (see also \cite{Loewer2020}).  In general, physics is non-linear and large effects of small changes are well known to happen.  From this perspective, agency is simply a situation where scale separation does not hold: nothing puzzling here. 
 
 What is puzzling, on the other hand, is why the branching is towards the future.  As discussed, since the microphysics is time reversal invariant, the reason for the time orientation of the branching can only be the time asymmetry of the macrophysics, namely the second principle.  How does this connection work?
 
Since choosing is irreversible it cannot happen without entropy increase.  Therefore during the interval $t_a-t_b$ there must be an entropy increase $\Delta S>0$.  On the other hand, suppose we observe the macroscopic evolution. Before the time $t_a$ we have no information about which branch will the system follow.  After the time $t_b$ we can see which branch has been realised, hence we have novel information. Where does the information come from?   The only possible answer is that $I$ is paid for by the increase in entropy $\Delta S$. 

A model illustrating how this can happen was developed in \cite{Rovelli2020} to account for the relation between memory and entropy. Let us adapt it here to the present case. 

Consider two systems: a system $\cal A$ (Agent) at temperature $T_a$ and a system $\cal W$ (World) at a lower temperature $T_w<T_a$. Assume that the two interact only occasionally, say on average once every $\cal T$ seconds, and weakly, namely with a long thermalization time $\tau_a \gg \cal T$. Furthermore, say that $\cal W$, in turn, is formed by $N$ subsystems, also interacting weakly among themselves, but with a global thermalization time $\tau_w\ll \cal T$.  Remarkably, these meagre ingredients are sufficient to model an agent.

From the definition of the thermalization time (on average $dT_a/dt=-T_a/\tau_a$) the average change of temperature $\delta  T_a$ during the interval $\cal T$, hence at each interaction, is given by 
\be
\frac{\delta T_a}{T_a}=-\frac{\cal T}{\tau_a}
\ee
Assuming for simplicity that the heat capacity of $\cal W$ is infinite and calling $C$ the heat capacity of $\cal A$, the average exchanged energy in one interaction is $Q=-C\delta T_a$, giving
\be
Q= CT_a \frac{\cal T}{\tau_a}.
\ee
This is heat, since it comes from the thermal energy of $\cal A$. Since  $\tau_w\ll\cal T$, in a typical configuration the $N$ subsystems of $\cal W$ have thermalised and have equal mean energy, say $E_i=E$, where $i=1,...,N$. We take the $N$ quantities $E_i$ to be macroscopic observables. With a frequency dictated by $\cal T$, the interaction between $\cal A$ and a random variable of  $\cal W$ happens.  Because of the second law, it is more likely that energy is transferred from $\cal A$ to $\cal W$ than viceversa.  On average, at each interaction an amount $Q$ of energy is transferred to one of the $N$ components of $\cal W$, say $i=\hat i$. After the interaction and for a time of the order of $\tau_w$, the energy of one of the $N$ components of $\cal W$ is higher than the others.  Therefore the macroscopic state of the system around an interaction happening at a time $t_o$ is described by 
\begin{eqnarray}
E_i(t)&=&E, \ {\rm for}\ t<t_o, \\
E_{i}(t)&=&E,  \ {\rm for\ } i\ne\hat i \ {\rm and\ }\ t>t_o,\\
E_{i}(t)&=&E+Q,  \ {\rm for\ } i=\hat i \ {\rm and\ }\ t>t_o,
\end{eqnarray}
which satisfies \eqref{agency}, and is therefore an example of agency. Thus, the simple thermodynamical ingredients above can give rise to a system that chooses and influences the \emph{future}. 

The interaction selects one out of $N$ alternative, producing an amount 
\be
I=log_2\ N
\ee
of information. The process is irreversible, because heat moves from a hot to a cold body, and produces an entropy increase.
\be
  \Delta S\sim \frac{Q}{T_w}-\frac{Q}{T_a}\sim   C \frac{\cal T}{\tau_a} \frac{\Delta T }{T_w}, 
\ee
where $\Delta T=T_a-T_w$.  A necessary condition for the information to be accounted for is 
\be
I<\Delta S, 
\ee
because information must come from somewhere. Using the equations above, this gives 
\be
N< 2^{\frac{\cal T}{\tau_a} \frac{C\Delta T }{T_w}}.
\label{bound}
\ee
This equation bounds the possibility of choosing between alternatives, at given thermodynamical parameters.  In particular, it shows that a non-vanishing temperature difference $\Delta T$ is needed to have a choice.  

To get a sense of this bound, consider it in a very simple case. Consider a minimal choice between 2 alternatives, namely $N=2$; using ${\cal T}\ll\tau_a$ we have
\be
C\Delta T \gg k\ T_w
\ee
where we have reinserted the Boltzmann constant $k\ne 1$ for clarity. The left hand side of this equation is the excess thermal energy in the agent, while the right hand side is the average energy per degree of freedom in the world.  That is: in order to be able to choose, the agents must have enough energy to stand up above the thermal energy of the world.

Crucially, there is no reduction of entropy in choosing: there is increase in entropy, contrary to what appears in the picture where the physics of the agent is disregarded.  Choosing is a conventional irreversible process, and it happens because it is statistically favoured, as all irreversible processes do.  

Before concluding, we comment on two points that we left open: top-down causation and the role of quantum theory.  

\subsection{Top down causation}

Defenders of top-down causation point out that it may be possible to account for the choice of an agent in terms of high-level concepts. For instance, a choice can be motivated by a value system, a calculation about the future, respecting a rule or a moral obligation, knowledge, memory, a computer program, or similar high-level notions.   This is obviously true, and does not alter the picture given above, for the following reason. 

High-level concepts make sense autonomously and permit us to predict events, but they nevertheless supervene on microphysics.   That is, two situations that differ in their high-level description cannot be identical in their microphysics.  For instance: it makes sense to understand the behaviour of a computer in terms of its software rather than thinking in terms of the forces on its elementary particles; but to have different software we necessarily need a different configuration in the elementary particles.  Equivalently:  it makes sense to understand the behaviour of a person in terms of her moral values, but to have different moral values must be accompanied by something different in the microphysics, perhaps in some synapses in the brain.  

Now, if high-level concepts are sufficient to account for behaviour, this is a normal case of causal closure of a coarse-grained account of the events.  High-level concepts, from this perspective, are normal macroscopic variables.  We are thus in the case of an internal agent, for which it \emph{is} possible to account for the choice: there is no entropy production in the choice, and the choice is fully determined by the macrophysics.   A computer playing chess, for instance, choses a move on the basis of rules.  This is an unproblematic case of causal closure of a macroscopic description. 

If, on the contrary, high level concepts are \emph{not} sufficient to account for behaviour, then we are back to the micro/macro context.  Something else is doing the choice: if it is not  the macrophysics, it must be  the microphysics. There are always very many micro-histories compatible with any given high-level account, leaving space for the branching.

Neither case conflicts with the causal closure of the microscopic physics.  Ultimately, agency is \emph{always}  nothing else than ignoring some physical links.   

\subsection{Quantum theory}

I have framed the discussion in terms of classical mechanics, because including probabilities complicates the language.  But nothing substantial changes in the above if quantum mechanics is taken into account.   

Microscopic time reversal invariance is not broken by quantum randomness \cite{Einstein1931,Rovelli2016}.  
The predictions of quantum mechanics are formulated in terms transition probabilities.  These do not distinguish between past and future and are time reversal invariant (CPT invariant in quantum field theory). The discussion in this paper, on the other hand, clarifies the origin of the time asymmetry in our conventional  \emph{use} of quantum theory. 
  We routinely interpret quantum transition probabilities as time oriented, namely we routinely read them as probabilities for future events given past events; but this is perspectival. It is because \emph{we} are agents that can influence the future, immersed in a time oriented macroscopic world, that we do so.   Therefore the time orientation of the common reading of quantum probabilities is just perspectival.   As shown, this perspectival time orientation, in turn, is ultimately sourced by the arrow of time of the second principle, via our own agency. 

Quantum theory does not change anything regarding the distinction between microphysics and macrophysics, either.   For the sake of the current discussion quantum indeterminism can be treated as due to unaccounted degrees of freedom.  If one wish to, one can even do so explicitly by using an interpretation of quantum theory like the de Broglie-Bohm hidden variable one, where indeterminism is indeed statistical ignorance, or Many Worlds, where indeterminism is indexical, namely ignorance of the branch in which we are located. Alternatively, one may simply remember that in order to affect the macro-world, quantum indeterminism needs decoherence, which is precisely based on disregarding degrees of freedom.  

Whether the causal closure of the macroscopic description of the world is in principle accounted by some underlying classical deterministic microphysics or by quantum randomness is irrelevant for  the understanding of agency.

\section{Conclusions: Memory and the creation of information}

We have a strong feeling that we cannot influence the past, but we can influence the future.  This seems  to conflict with the time (CPT) reversal invariance of fundamental physics. But is not.  We have this feeling because truly we can affect the macroscopic future but not the macroscopic past.  The macroscopic world we work with has a fixed past determined by abundant present traces and memories  \cite{Rovelli2020}, while it is compatible with a number of different futures, that do depend on what happens in our brain.    This is the openness of the future that our feeling veridically captures.  

This openness of the future leaves ample space for subtle high level processes to influence the macroscopic future.  Our sense of being free to decide is clearly rooted here.  It is in this sense, that, as Ismael puts it: "Physics Makes Us Free" \cite{Ismael2016}. 

The microscopic account is a wholly different story, but is of little relevance for our experience and feelings, since, by definition, we do not access it.  

Independent agency is a description of the \emph{macroscopic} dynamics of an interaction between an agent and the world which: (i) is unpredictable, (ii) is irreversible, (iii) produces a (macroscopically) detectable effect on the world in the future, and (iv) produces information.  There are remarkable similarities between this and the model for traces, or "memories",  described in  \cite{Rovelli2020}. 

Both memory and agency are events that leave a trace in the macroscopic domain.  The difference is that the roles are in a certain sense exchanged:  in the case of agency, it is the agent that leaves a trace on the external macroscopic world; while memory is a trace left by the world on the memory system. Both phenomena need long thermalization times, namely quasi-stable system, to hold the memory or the effect of the action.   Both need a disequilibrium in the past, to account for orientation and irreversibility.     Both can be understood as macroscopic phenomena pertaining to a coarse grained picture of the world, and make no sense at the microscopic level (except in metaphorical "anthropocentric" language). 

Agency is time oriented because it is a macroscopic phenomenon driven by an entropy gradient (hence ultimately by the past hypothesis). The model presented in Section \ref{Sec:4} shows that system separation, 
past temperature difference, and long thermalization times are meagre elements nevertheless sufficient to model this thermodynamical roots of agency. In turn, our time orientation as agents compels us to look at mechanical interactions in a time oriented manner \cite{Price}.

The most interesting aspects of the two phenomena is that they both produce information.  In agency, information can be recognised, in Shannon's sense, as the instantiation of one among a number of possibilities.  
In the case of memory, information is what Shannon calls "relative information": physical correlation  between a past macroscopic event and its trace. 

In both cases, the information is generated by increasing entropy. Eq \eqref{bound} gives the maximal information 
that can be produced in choosing, at given thermodynamical parameters.  It analogous to bound on the information produced by the formation of memory derived in  \cite{Rovelli2020}. 

 Low entropy is a form of information because a lower entropy state amounts to a more selective information about the microphysics (a zero entropy macrostate is a state that has maximal information about the microphysics: the microstate is unique).  Memory and agency utilise the information stored in low entropy and translate it into information readable in the macroscopic world.  In fact, they both can be viewed as mechanisms that generate macroscopic information. 

Macroscopic information, stored in human memory, in DNA molecules, in computer messages, in books, in narratives, in software codes, in records of any form, must have been ultimately produced by physical mechanisms.  Traces of the past and decisions by agents ---possibly in turn themselves affected by memories of the past--- are major sources of everything we call information.  In both cases, information is created, in a statistically favoured manner,  at the expenses of low entropy, in accordance with the second principle.   In a fully thermalised situation, there is no space for memories or for agents.   

The entire informational universe formed by the biosphere  and by culture can therefore perhaps be viewed, from this perspective, as formed by information produced by a mechanism of the form described here. \\

 \centerline{***}

I am deeply indebted with Jenann Ismael: several ideas of this paper developed in conversations with her.  Thanks to David Albert for a very insightful criticisms to an early draft of this paper, crucial for me.   This work was made possible through the support of the FQXi  Grant  FQXi-RFP-1818 and of the ID\# 61466 grant from the John Templeton Foundation, as part of the \emph{The Quantum Information Structure of Spacetime} (QISS) Project (\href{qiss.fr}{qiss.fr}).


\onecolumngrid

\providecommand{\href}[2]{#2}\begingroup\raggedright\endgroup

\end{document}